\documentclass[prb,aps,showpacs,superscriptaddress,floatfix,preprint]{revtex4}	

\usepackage{graphicx}
\usepackage{dcolumn}
\usepackage{bm}
\usepackage{amssymb,amsmath}
\usepackage[usenames]{color}

\begin{document}


\title{Mott transition in the Hubbard model away from particle-hole symmetry}

\author{D. J. Garc\'{\i}a}
\affiliation{Instituto de F\'{\i}sica Gleb Wataghin, Unicamp, CEP 13083-970
Campinas, SP, Brazil}
\author{E. Miranda}
\affiliation{Instituto de F\'{\i}sica Gleb Wataghin, Unicamp, CEP 13083-970
Campinas, SP, Brazil}
\author{K. Hallberg}
\affiliation{Instituto Balseiro and Centro At\'omico Bariloche, CNEA, (8400) San 
Carlos de Bariloche, Argentina}
\author{M. J. Rozenberg}
\affiliation{Laboratoire de Physique des Solides, CNRS-UMR8502, Universit\'e de 
Paris-Sud, Orsay 91405, France.}
\affiliation{Departamento de F\'{\i}sica, FCEN, Universidad de Buenos Aires, 
Ciudad Universitaria, Pabell\'on 1, Buenos Aires (1428), Argentina.}

\date{\today}

\begin{abstract}
We solve the Dynamical Mean Field Theory equations for the Hubbard model away from the particle-hole symmetric case
using the Density Matrix Renormalization Group method.
We focus our study on the region of strong interactions and finite doping where two solutions coexist.
We obtain precise predictions for the boundaries of the coexistence region. In addition,
we demonstrate the capabilities of this 
precise method by obtaining the frequency dependent
optical conductivity spectra.
\end{abstract}

\maketitle

{\it Introduction}: Solving models for strongly correlated compounds (typically those 
including {\it d} or {\it f} electrons)  remains a hard and
important open problem of modern condensed matter physics. 
In these systems interesting anomalous behavior usually occurs as a consequence
of the  competition between the kinetic and Coulomb energy of electrons, which are
of the same order of magnitude, 
and analytical methods based on perturbative considerations are notoriously unreliable.
Therefore, one is led to resort to non-perturbative techniques and numerical methods 
to try to deal with these difficulties.

In recent years, a non-perturbative approach to general strongly
correlated electron models, termed Dynamical Mean Field Theory (DMFT),\cite{review,pt} allowed progress in the understanding of several
physical problems, including the correlation-driven Mott
metal-insulator transition.  The key feature of DMFT is that it maps
the original lattice problem onto a self-consistent quantum impurity
model. This resulting quantum impurity remains, nevertheless, a fully
interacting many-body problem that has to be solved.\cite{review} The
success of DMFT in dealing with model Hamiltonians has generated a
great deal of interest in combining it with \textit{ab initio}
band structure methods with the goal of obtaining a realistic
description of correlated electron compounds.\cite{pt}  The main
technical difficulty is the lack of a reliable method to solve the
associated quantum impurity problem. Currently, the most widely
adopted methodology is the quantum Monte Carlo technique, which
generally requires an analytical continuation of the results to the
real frequency axis. Unfortunately, the latter introduces some
uncertainty in the procedure, which becomes a severe problem for
multi-orbital systems. Therefore, there is strong interest in the
development of new methods that can deal with general quantum impurity
models directly on the real axis. An interesting proposal was recently
introduced which was based on the precise diagonalization of the
quantum impurity Hamiltonian with the powerful Density Matrix
Renormalization Group (DMRG).\cite{garkarmar,nishimoto,KarskiRaasUhrig} This
technique has the virtue of being already widely employed in the study
of low-dimensional strongly correlated systems and is based on a
judicious trimming of the Hilbert space so as to restrict the
numerical calculation to its most relevant subspace.\cite{dmrg} In
this manner, despite the exponential growth of the Hilbert space,
hundreds of orbitals can be taken into account.  Moreover, the method
is not restricted to the ground state and finite energy excitations
can be computed with similar accuracy.\cite{Karen} This is in
contrast to the related alternative, the Numerical Renormalization
Group (NRG) method, which aims at a very precise description of
the low frequency quasiparticle peaks associated with low-energy
excitations, as done in its celebrated application to the Kondo
problem.\cite{Wilson} The DMRG, on the other hand, treats high and low
frequencies on an equal footing.

The initial DMRG solution of the DMFT equations was obtained for the
simplest test case of the particle-hole symmetric Hubbard model.\cite{garkarmar,nishimoto,KarskiRaasUhrig} In this paper we show that the method can
reliably tackle the general finite doping case. In particular we focus
on the most demanding region of the phase diagram where two 
solutions coexist near the correlation-driven Mott metal-insulator
transition and obtain the phase boundaries with unprecedented
precision. We also illustrate the capabilities of the methodology by
computing the frequency-dependent optical conductivity, which requires
the reliable description of higher energy features, such as the
Hubbard bands, that lie beyond the scope of the NRG method. Our
results show that even at the smallest dopings and strong interaction
strength, the low frequency contribution to the optical conductivity,
the Drude part, remains very well described by a Lorentzian
line-shape. We also show that both the vanishing of the Drude weight
or the doping can signal equally well the destruction of the
correlated Fermi liquid metallic state.  Our work amounts to a
substantial improvement with respect to previous studies based on
exact diagonalization (of up to 8 sites, however nowadays somewhat
larger systems are possible with ED) and an {\it ad hoc}
perturbation scheme,\cite{kajueter} and opens the way for the
application of the DMFT+DMRG method to more general model
Hamiltonians. Details of the method will be given elsewhere\cite{preparation}.

{\it The Model:}The Hamiltonian of the Hubbard model is defined by
\begin{equation}
H=\frac{t}{\sqrt{2d}}\sum_{\langle i,j\rangle, \sigma} c_{i,\sigma}^{\dagger} c_{j,\sigma} + U\sum_i (n_{i,\uparrow}-\frac{1}{2}) (n_{i,\downarrow}-\frac{1}{2}) - \mu \sum_{i,\sigma} n_{i,\sigma}
\label{hubbard}
\end{equation}
where $U$ is the on-site Coulomb interaction, $\mu$ is the chemical potential, $t$ is the hopping and $d$ is the space dimension.
We take the half bandwidth of the non interacting model as unit of
energy, thus $D=2t=1$. We particularize to the case of the
infinite-dimensional Bethe lattice, in which the non-interacting
density of states (DOS) is $D(\epsilon)=(2/\pi)\sqrt{1-\epsilon^2}$.

The associated impurity problem is the single-impurity Anderson model
(SIAM) and its hybridization function needs to be self-consistently
determined.\cite{review}  The SIAM Hamiltonian is solved in linear
chains\cite{qimiao,mpl} using the DMRG algorithm\cite{garkarmar} of up
to 101 sites keeping 128 states per block.

\begin{figure}[htb]
\includegraphics[width=\columnwidth,clip]{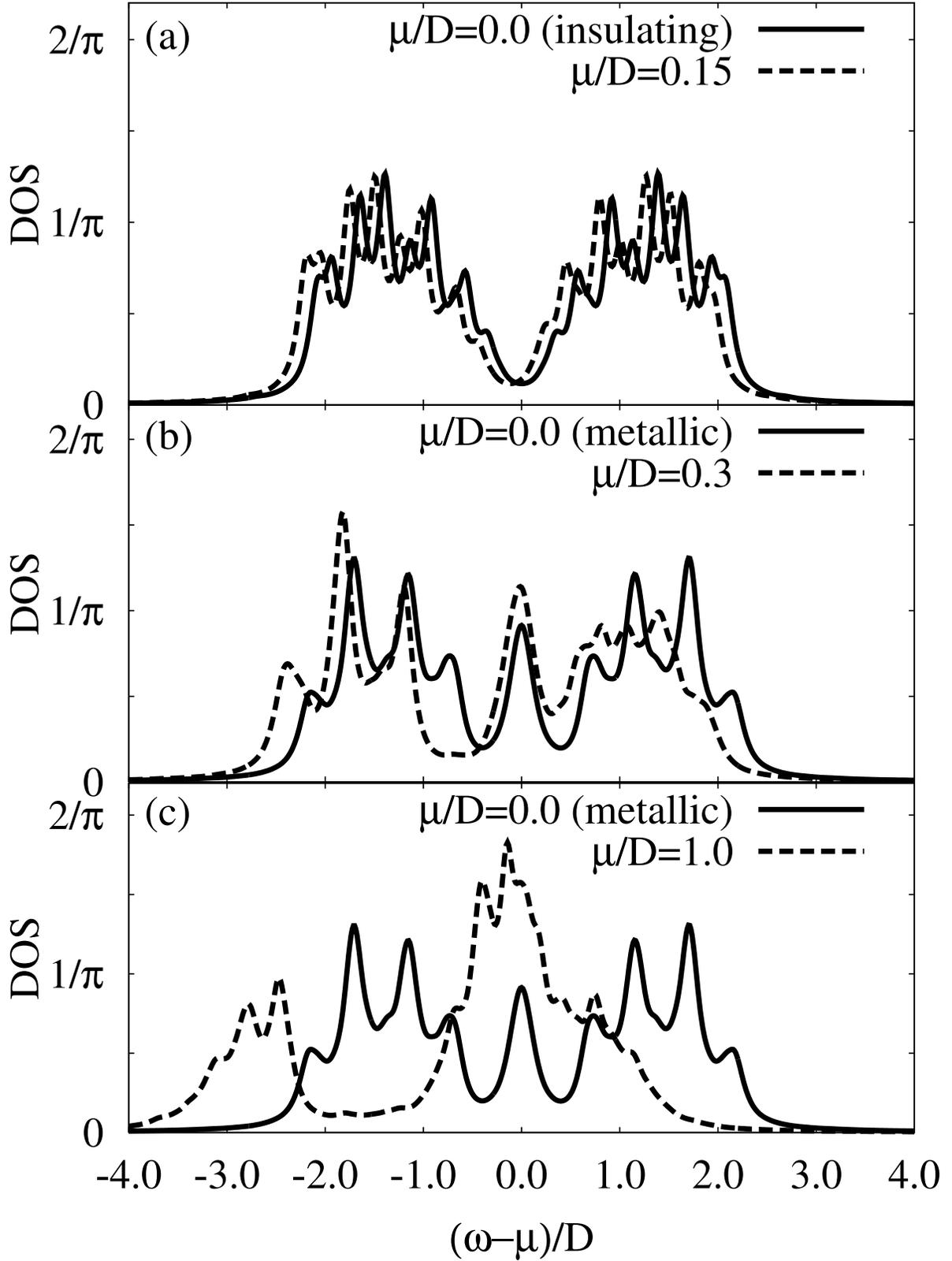}
\caption{
(a) Insulating, (b) lightly doped and (c) heavily doped metallic densities of states (DOS) 
of the Hubbard model for $U/D=2.6$. In (a), the small DOS weight seen at $\omega=0$ is due to the relatively large relation between the gap for $U/D=2.6$ ($\sim 0.19$) and the small Lorentzian broadening used($\eta=0.1$). 
\label{DOSvsmu}} 
\end{figure}

We focus our study on the poorly explored region of the $U$ - $\mu$
parameter space where two solutions, insulating and metallic, can be
obtained from the DMFT equations. In Fig. \ref{DOSvsmu} we show the
evolution of the DOS for the two solutions as one moves away from the
half-filled particle-hole symmetric case. The chemical potential $\mu$
is increased at fixed $U$.  While the use of DMRG allows the precise
diagonalization of the associated SIAM problem with a bath that can be
accurately described using around one hundred sites, this impurity
model remains, nevertheless, finite. Therefore, the computed Green's
functions contain a finite, though large, number of poles. Thus, as in
any exact diagonalization scheme, one needs to broaden the poles to
allow the observation of the DOS structure on the real axis. 

{\it Results:} We show in Fig. \ref{DOSvsmu} the results obtained by using a simple Lorentzian broadening $\eta$. We find that choice to be more
appropriate than the ``logarithmic'' broadening usually adopted in NRG
calculations\cite{Bulla,BullaPRB} which, although capable of sharply
resolving the insulating gap, tends to wash out the high energy
features of the density of states.  The results show that, in the
insulating case, when the chemical potential is moved within the Mott
gap, the lower and upper Hubbard bands shift rigidly, without any
ostensible transfer of spectral weight taking place
(Fig.~\ref{DOSvsmu}(a)).
The apparent substructure in the Hubbard bands seen in the insulating
DOS results from finite-size effects\cite{aclaracion}, i.e., a finite number of poles. 
Our finite size analysis (not shown) suggests that in the infinite chain limit 
the Hubbard bands become smooth. This is in contrast to what is seen in the large $U/D$ antiferromagnetic phase.\cite{sangiovanni}
In the lightly doped case one				 
observes that, as the central quasiparticle peak rapidly moves through
the region between the Hubbard bands, there is a transfer of spectral
weight as well as an evolution of the line shapes
(Fig.~\ref{DOSvsmu}(b)).  More precisely, one finds that the
quasiparticle peak receives spectral weight from both Hubbard
bands. For larger values of $\mu$, as the system gets heavily doped,
one finds that the quasiparticle peak eventually broadens as it merges
with the closest Hubbard band (Fig.~\ref{DOSvsmu}(c)).  As these
features coalesce, 
 they also draw spectral weight from the other
Hubbard band that remains at an energy distance of the order of
$U$.\cite{review}

\begin{figure}[htb]
\includegraphics[width=\columnwidth,clip]{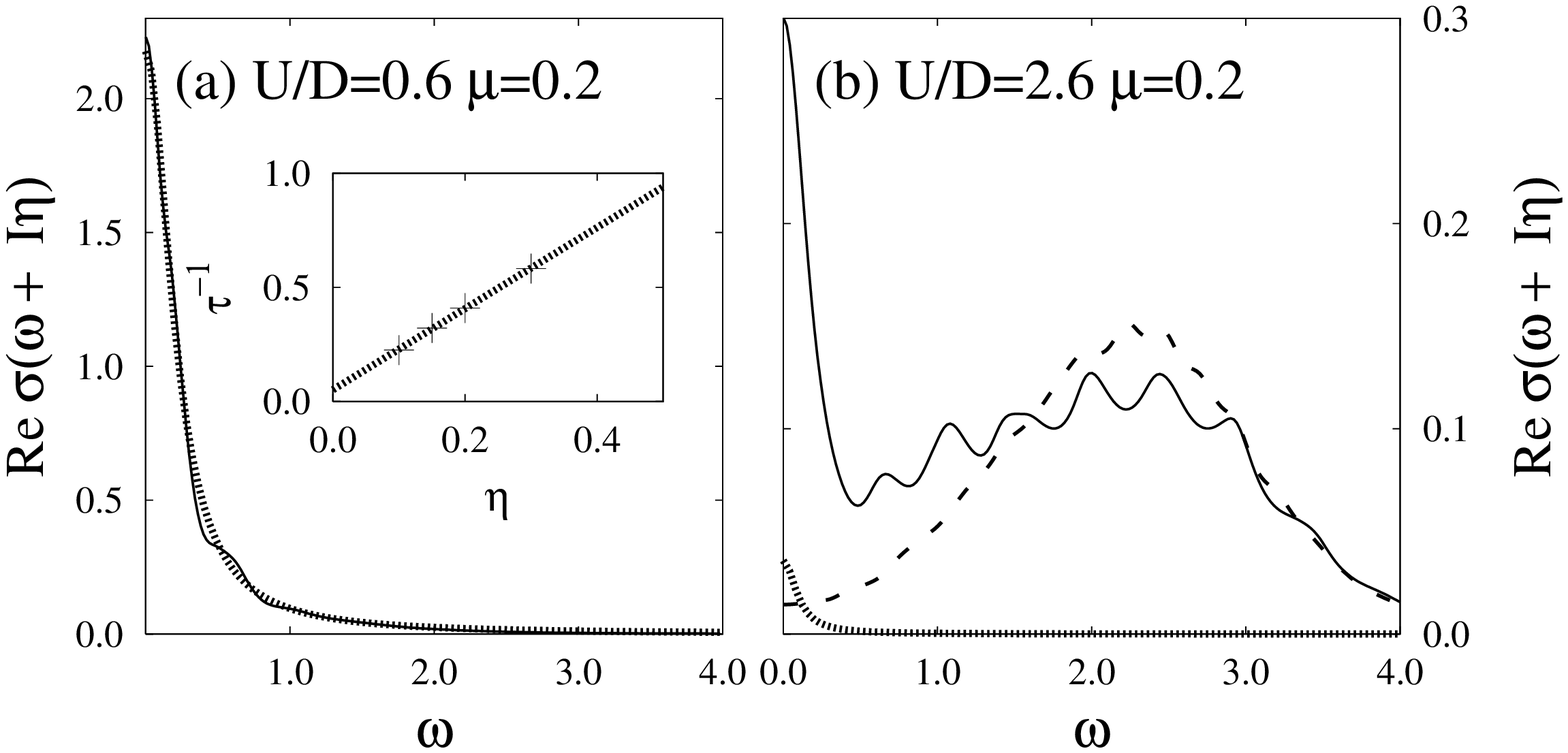}
\caption{Metallic (solid line) and insulating (dashed) optical conductivities for 
the Hubbard model in the purely metallic (a) and coexistent (b) regimes. The dotted 
line is a Lorentzian low-frequency fit (Drude model). For small $U/D$, the data and the 
Lorentzian fit agree in almost all the frequency range. The broad feature in (b) 
corresponds to the regular part explained in the text.
A small imaginary part $\eta=0.1$ has been used. 
The inset shows the evolution of $\tau$ in the Lorentzian fit of the low 
frequency conductivity as a function of $\eta$ for $U/D=0.6$.
\label{OptCond}} 
\end{figure}

In order to demonstrate the capabilities of the method we shall now compute the frequency-dependent optical
conductivity.
From the lattice Green's $G(\epsilon_{\bm{k}},\nu)$ function, where $\epsilon_{\bm{k}}$
is the non-interaction dispersion, we can
evaluate the optical conductivity within DMFT as\cite{review,RKK}
\begin{eqnarray}
Re~\sigma(\omega+I 0^{+})&=& \frac{\pi e^2}{\hbar a d}
\int_{\infty}^{\infty} d \epsilon D(\epsilon)
\int_{\infty}^{\infty} d \nu \\ \nonumber
&& \rho(\epsilon,\nu)\rho(\epsilon,\nu+\omega) \frac{\theta(\nu+\omega)-\theta(\nu)}{\omega}
\end{eqnarray}
where {\it a} is the lattice spacing, {\it d} is the spatial
dimension, $\rho(\epsilon,\nu)=\mathrm{Im}G(\epsilon,\nu-I 0^+)/\pi$,
and $I 0^+$ denotes an infinitesimal imaginary part.
For simplicity we have chosen to use an unitary vertex.
The evaluation of $\rho(\epsilon,\nu)$ 
requires the previous computation of the local self-energy. While in the standard exact diagonalization
solution of the DMFT equations this is a cumbersome procedure due to the small number of Green's function poles, the use of DMRG dramatically changes the situation and reliable $\Sigma(\omega)$ on the real
axis can be easily obtained from the self-consistency condition.\cite{review}
In Fig.~\ref{OptCond} we show the optical conductivity for two coexistent solutions (for parameters 
$U/D=2.6$ and $\mu=0.2$) and for the metallic state for weak interaction ($U/D=0.6$).
In the metallic case we see that, despite the very small doping, the small frequency regime of
$\sigma(\omega)$
can be very well described by a simple Lorentzian form that follows
from a Drude model\cite{Ziman}
\begin{equation}
Re~\sigma(\omega+I 0^{+})=\frac{DW \tau}{1+(\omega\tau)^2}
\end{equation}
where $\tau$ is the relaxation time and $DW$ is the Drude weight,
which is a measure of the number of quasiparticle carriers in the metal.\cite{DWclassical}
In our metallic case the finite value of $\tau^{-1}$ comes from the
finite imaginary part used to compute the Green's function $G$.
The inset shows that as $\eta$ goes to zero, $\tau^{-1}$ tends to zero
and we recover the delta-function behavior that corresponds to a clean Fermi Liquid.
For large $U/D$ we observe that, in addition to the small Drude part, the optical conductivity spectrum
has a large mid-infrared contribution at frequencies of order $U$.
This regular part corresponds to finite frequency optical excitations between the two Hubbard bands
and between the latter and the central quasiparticle peak and is almost absent for small values of the Coulomb interaction.

\begin{figure}[htb]
\includegraphics[width=\columnwidth,clip]{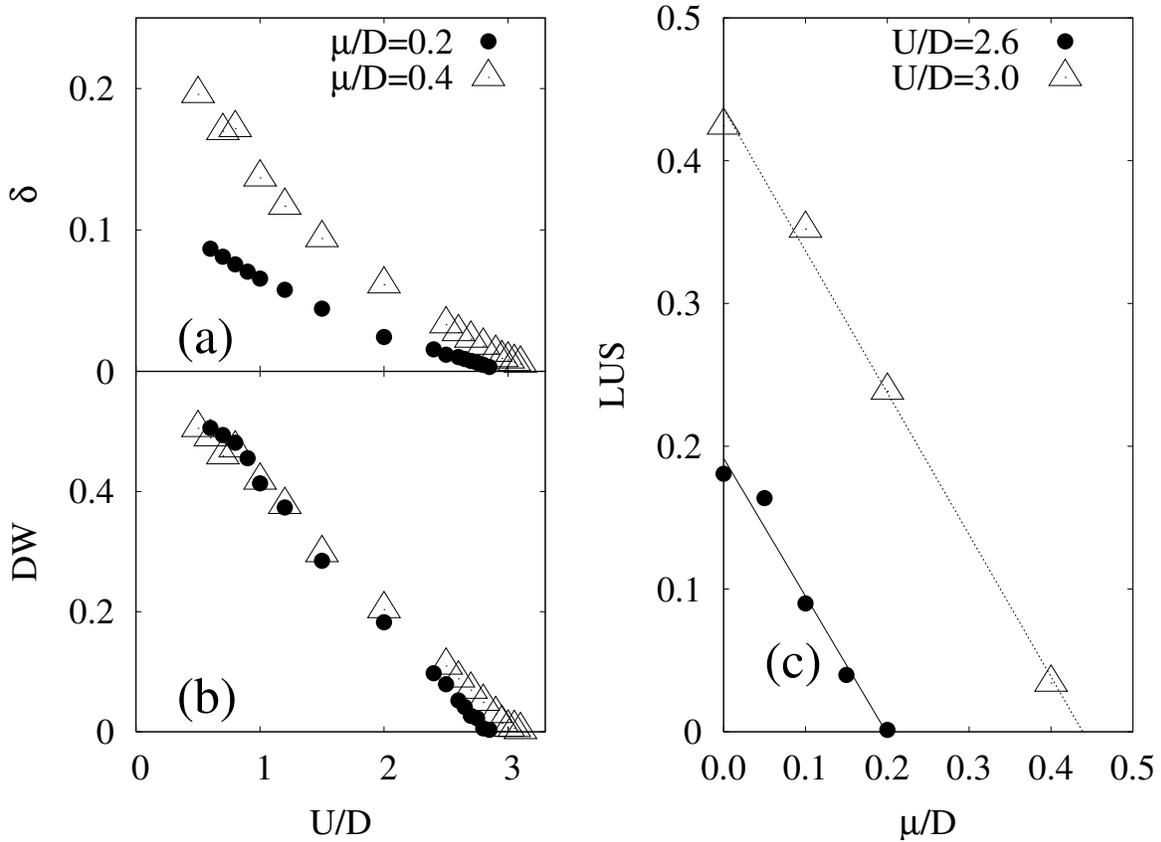}
\caption{(a) Doping $\delta$ and (b) Drude Weight DW for the metallic states for various $U/D$ and $\mu/D$ values.
(c) Energy of the lowest unoccupied state (LUS) in the upper Hubbard band for insulating solutions. \label{critval}} 
\end{figure}

The destruction of a normal metallic state is formally defined by the vanishing of the Drude weight, which signals the
localization of all metallic conduction carriers.\cite{kohn} On the other hand, in the Hubbard model it is well known
that, unless in phases with some type of long range order or in the presence of disorder, infinitesimal doping 
at $T=0$ is enough to drive the system
across a density driven metal-insulator transition. Therefore, a demanding test for the present method would be
to verify that both the doping $\delta=(n-1)/2$ 
(where $n$ is the number of particles per site)
and the Drude weight (DW) vanish at the same value of interaction strength.
We emphasize the important technical fact that, while the value of the doping is simply and accurately
computed from the expectation value of the number
operator, the Drude weight is, in contrast, independently obtained via the comparatively far more
laborious procedure described above.
The comparison of the results of panels (a) and (b) of Fig.~\ref{critval} shows that, in fact, 
both observables, $\delta$ and DW, are found to vanish at
equal interaction values for each choice of the chemical potential.
 
Fig. \ref{critval}(a) shows the evolution of the doping for fixed chemical potential, varying the Coulomb interaction.
The doping increases as $\mu$ moves to larger values, i.e., away from the particle-hole case.
At fixed $\mu$, increasing the correlation $U$ from the
non-interacting limit acts to decrease $\delta$
continuously to $0$, where the metallic solution is no longer stable
and gives rise to the insulating one.
The extrapolation of the lowest doping values towards zero for different chemical potentials provides an accurate 
estimate of the critical line $\mu_{c2}(U)$ which locates  the instability of the metal towards an insulating solution.
The Drude Weight DW is shown in Fig. \ref{critval}(b). Its behavior is qualitatively different from that of $\delta$, 
since $DW(U)$ does not uniformly increase with increasing $\mu$. 
In the low $U/D$ region, the DW decreases as the chemical potential is increased, reflecting the lowering of the kinetic 
energy due to the fewer number of carriers. In contrast,
for larger values of the interaction close to the $\mu_{c2}$ line, the DW decreases as $\mu$ decreases towards
particle-hole symmetry, reflecting the enhancement of the effective mass as the metal-insulator
transition is approached. 

\begin{figure}[htb]
\includegraphics[width=\columnwidth,clip]{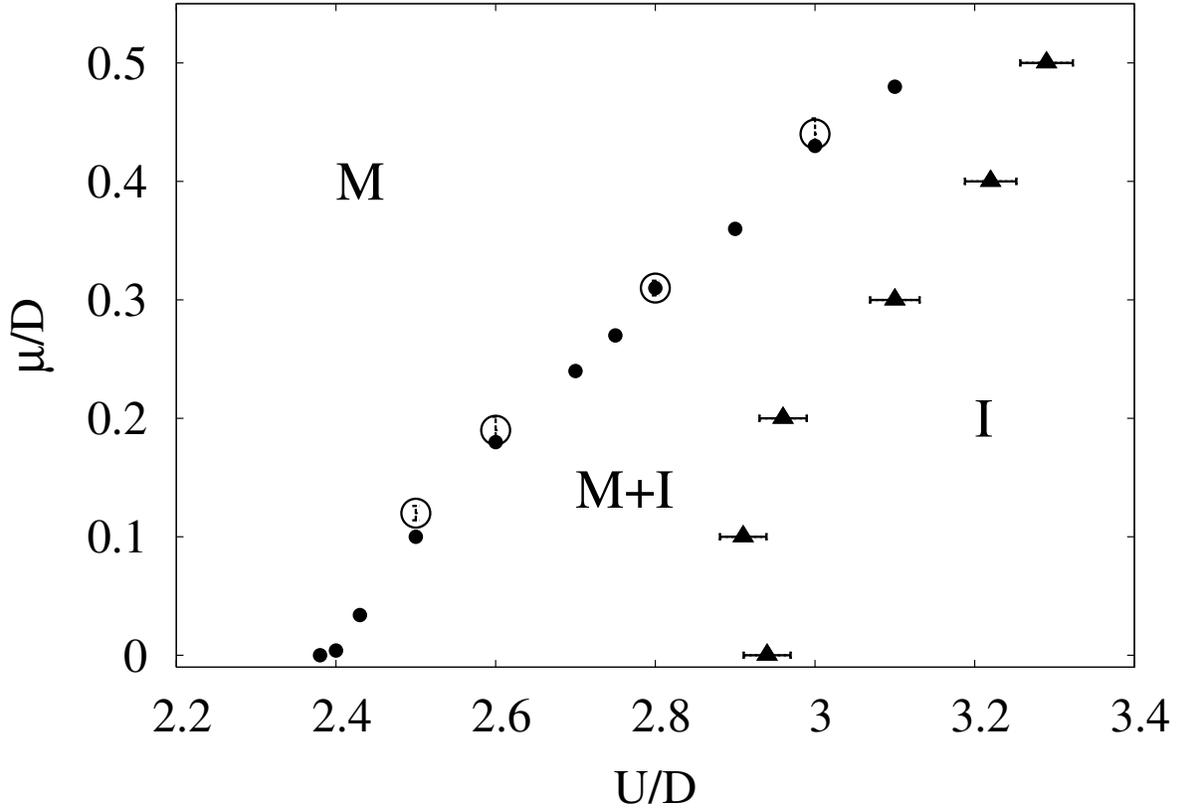}
\caption{Phase Diagram: Small full circles show half the gap value for $\mu=0$. Empty circles (full triangles) are the extrapolated critical values of the chemical potential $\mu_{c1}$ ($\mu_{c2}$) where the insulating (metallic) solution no longer exists.
\label{diagfases}} 
\end{figure}

It is also possible to investigate the instability of the insulating state towards the metal. This transition
is signaled by the collapse of the Mott-Hubbard gap as the chemical potential is brought to a Hubbard band edge.
Following the energy of the lowest unoccupied state (LUS) in the upper
Hubbard band with respect to the Fermi 
level in the insulator as $\mu$ increases, it is possible to determine the transition line 
$\mu_{c1} (U)$ (Fig.~\ref{critval}(c)) as the value of the critical chemical potential for which 
the energy of the LUS vanishes.
As the bands move in an approximately rigid way for $\mu < \mu_{c1} (U)$ the value of the chemical 
potential varies linearly and agrees with half the size of the band
gap at $\mu=0$ (see Fig.~\ref{diagfases}).
For $\mu>\mu_{c1}(U)$ a finite number of poles appears at positive and negative small values of 
$\omega-\mu$, signaling the metallic state.

The critical lines allow us to accurately draw the phase diagram of
the model away from particle-hole symmetry at $T=0$. Previous work has
been restricted to either a small number of sites\cite{kajueter} in
the effective impurity Hamiltonian or to small but finite temperatures
using quantum Monte Carlo.\cite{MKR} On the other hand, the NRG
method is not very well suited for the accurate investigation of the
insulating state.  In Fig. \ref{diagfases} we plot the $\mu_{c1}(U)$
and $\mu_{c2}(U)$ lines that determine three regions in the $\mu$-$U$
phase diagram: For $\mu > \mu_{c1}(U)$ ($\mu < \mu_{c2}(U))$ only
metallic (insulating) solutions are found.  
In the middle there is a region of coexistence of both kinds of states,
 where the metallic state is one with the lowest energy.\cite{review}
The phase diagram presented here shows an overall agreement with the one obtained
through exact diagonalization in the ``star geometry''\cite{kajueter}
where the impurity site is connected with hopping terms to all the
other sites. The main differences are found for the $\mu_{c2}(U)$ line
because, as the metal to insulator transition is approached, the
quasiparticles develop a diverging mass corresponding to a very narrow
quasiparticle peak. In the language of the associated SIAM, this
narrow resonance implies a large correlation length which can be fully
realized only in long enough systems. This can only be obtained with
the method presented here, allowing for very accurate results.

{\it Conclusions:} In this work we have shown that the DMRG method, in addition to 
being
largely used to compute spectral quantities of low dimensional
strongly correlated systems,\cite{Karen,dmrg} allows for a practical
implementation of an accurate impurity solver of the DMFT equations of
the Hubbard model in a general case. We have computed spectral
functions including the DOS and the frequency-dependent optical
conductivity. We have also calculated the behavior of the doping and
the Drude weight as a function of the chemical potential near the
metal-insulator transition and demonstrated the accuracy of the method
by passing the demanding test of the comparison of their respective
predictions for the metal-insulator critical line. Due to the fact
that with this method long enough systems can be handled, these
critical lines can be very accurately obtained.
Of course, in the non-frustrated case the true ground state of the model is AF at
 low doping and the extension of the method to that case is left for future work.

The implementation of the DMRG method for the Hubbard model in the
non-symmetric case is an important step towards achieving an exact,
unbiased and general impurity solver to be used in the realistic {\it
ab initio} strongly correlated electronic structure calculation
program.\cite{pt} The next step ahead is to generalize the
methodology for the multi-orbital case, where interesting physical
problems remain open, such as the orbital-selective Mott transition
with a fully rotationally invariant Hamiltonian.

DJG acknowledges support form the FAPESP and CAPES-BA programs.  EM
acknowledges support from FAPESP and CNPq. Part
of the computations were performed with the aid of CENAPAD-SP (National
Center of High Performance Processing in S$\tilde{\rm a}$o Paulo), project UNICAMP /
FINEP - MCT.  We also acknowledge support from CONICET, the Guggenheim
Foundation, PICT 03-06343 and 03-13829 of ANPCyT.


\begin{thebibliography}{20}
\bibitem{aclaracion} In Ref. \onlinecite{nishimoto} the  algorithm used includes a spectral ``smoothing" which consists on averaging the spectral weight within each energy interval and where the self-consistency is searched within this solution. In our case, instead, we do not average the spectral weight and the peaks obtained correspond to the finite size chain considered.
\bibitem{DWclassical} In a classical approximation, $DW=\frac{n e^2}{m}$ where $n$ is the carrier density and $m$ is its mass.
\end{thebibliography}
\end{document}